# *Eurythmic Dancing with Plants*
## Measuring Plant Response to Human Body Movement in an Anthroposophic Environment

Sebastian Dürr[1], Josephine van Delden[1,2], Buenyamin Oezkaya[1,3], Peter A. Gloor[1]
[1]MIT, [2]TU Munich, [3]RWTH Aachen
contact: pgloor@mit.edu

## Abstract

This paper describes three experiments measuring interaction of humans with garden plants. In particular, body movement of a human conducting eurythmic dances near the plants (beetroots, tomatoes, lettuce) is correlated with the action potential measured by a plant SpikerBox, a device measuring the electrical activity of plants, and the leaf movement of the plant, tracked with a camera. The first experiment shows that our measurement system captures external stimuli identically for different plants, validating the measurement system. The second experiment illustrates that the plants' response is correlated to the movements of the dancer. The third experiment indicates that plants that have been exposed for multiple weeks to eurythmic dancing might respond differently to plants which are exposed for the first time to eurythmic dancing.

## 1. Introduction

In late 1900, researcher Victor Schauberger observed a farmer who used "an ancient Alpine tradition" (Bartholomew 2003), "Tonsingen", German for "melodic singing" (Coats 1996) to stimulate growth of his plants. The quality of the crops in the farmer's fields was much higher compared to his neighbors' fields (Coats/Callum 1996). This farmer used methods of fertilizing his fields by singing to them comparable to methods employed today by biodynamic farmers (Joly 2005, Proctor 1997).

In a biodynamic agricultural environment, Stucki (2010), Retallack (1973), Ekici et al. (2007), Singh (1962) as well as Wachsmuth (1932) and Poppelbaum (1952), focusing on the the anthroposophic aspects, assert that human stimulation (e.g. eurythmy) positively affect plants and stimulate their growth.

However, *experiments to measure these anthroposophic effects* have not been conducted in their research. To close this gap, in this paper we investigate the impact of eurythmic dances on plants by measuring leaf movement and variations in plant electrical discharge in an anthroposophic experiment at a biodynamic research facility.

Eurythmy is defined as making visible language and music through "regularities and relationships through human movement" (Buessing et al. 2008). Biodynamic agriculture (Proctor 1997) is an alternative approach to farming with the goal to balance agriculture and the environment with methods including *etheric energies* (Steiner 1924, Guzzon et al. 2016).

We build on scientific measurement practices by Bose (1919), and Oezkaya and Gloor (2020), who found leaf movement and electrical discharge as a primary response mechanisms of plants.

By addressing the research question *"How do plants respond when being exposed to eurythmic dancing in terms of electrical discharge and movement?"*, we empirically measure plants' response behavior in their leaf movement, and electrical discharge to eurythmic stimulations.

Our results reveal significant correlations between the eurythmy dance and the plant's movement pattern, as well as its electrical discharge, which anthroposophic researchers would deem a reaction to "etheric energies", i.e., the lowest layer in the "human energy field" or "aura" (Jacka 2011).

In the following sections, we review the literature on anthroposophy, biodynamic agriculture, eurythmy, and plant responses (movement, electrical discharge), introduce and analyze the experiments and then discuss the main findings and limitations.

## 2. Theoretical Background

### Anthroposophy and Biodynamic Agriculture

Anthroposophical studies examine the living energies apparent in plant life, and describe them in terms of etheric activity (Steiner 1924, Wachsmuth 1932, Bockemühl 1977, Keats 1999, Proctor 1997). In the anthroposophical perspective, the energetic process of air being set into wave-like motion (Wachsmuth 1932) can be achieved by applying eurythmy (Kirchner-Bockholt 1977). Wachsmuth (1932) attributes this effect to the interplay of contrary etheric forces.

Biodynamic agriculture is an anthroposophic approach to farming that originates from Rudolf Steiner (1924). The goal is to balance farming and the environment through preventing animal and plant diseases. The method to achieve this includes spiritual, ethical and ecological elements (Wachsmuth 1932, Grohmann 1968). Thus, biodynamic farming is distinguished from organic farming as it is said to reject modern farming approaches (e.g., refusing strong usage of artificial fertilizers) (Philipps et al. 2006).

### Eurythmy

Eurythmy is an artistic movement that was developed from Rudolf Steiner's insights (Usher/Beth 2006) at the beginning of the 19th century, becoming a distinctive feature in the practice of anthroposophy (Bockemühl 1977). The word "eurythmy" consists of the Greek word 'eu' which means harmonious or beautiful and 'rhythm' (from the Greek word 'ruthmos') which can be translated with 'flow'. Eurythmy conveys either spoken or musical sounds through choreography and gestures (Usher/Beth 2006). Eurythmy consists of two parts: Firstly, there is speech-eurythmy, i.e. the movements and gestures to embody and unveil spoken sound (Kirchner-Bockholt 1977). Secondly, there is tone-eurythmy, i.e. any gestures and movements which aim at embodying and producing visible musical sound (Steiner 1924). Richard's (1995) work shows that certain speech-eurythmy gestures affect plant movement.

## Plant Movement and Electrical Discharge

Bose (1919) pioneered the field of electrical signaling of plants, where he emphasized bioelectric potentials and movements (Johnsson et al. 2012). Working mainly with Mimosa Pudica and Desmodium Gyrans (alternatively named "Codariocalyx motorius"), Bose measured the mechanical and electrical responses of these plants to external stimuli. A reaction of these plants was found for a broad range of stimuli including electrical charge, mechanical load, and temperature (Johnsson et al. 2012).

Plants developed pathways for electrical signal transmission to respond rapidly when external stimuli are applied (Volkov 2006). For example, floral electric fields can be impacted by potential differences between flowers and insects (Clarke et al. 2013). An electrical response can also be triggered as a result of an injury to the plant. An injured leaf starts producing an electrical signal that then spreads to adjacent leaves (Christmann, Grill 2013). The signals are best transmitted to those parts of the plant that have strong connections through the vasculature system. Christmann and Grill (2013) suggest that these electrical signals and their creation and dispersion are an essential part of a plant's reaction to defend itself against herbivores.

Next to electrical discharge, movements can be a reaction of plants to external stimuli. The research on plant movement goes back to Charles and Francis Darwin's publishing "The Power of Movement in Plants" (1880). Plant movements can be distinguished into tropisms which are the "directed growth in response to external stimuli" and nastic movements, which are a "response to stimuli, but the direction is independent relative to the stimulus source" (Kiss 2006).

In line with this theoretical background, anthroposophical research (e.g., Wachsmuth 1932, Steiner 1924) contends that human stimulation (e.g. eurythmy) affects plants and influences their movement. Next, we quantify these effects by measuring the nastic movements and electrical discharge as suggested by Bose (1919).

## 3. Method

Following Bose (1919), we study the plants' electrical discharge to measure the effect of eurythmy on plants. Additionally, Steiner (1904,1994) states that a living plant, comprising anthroposophical terms both a physical and an energetic body, is constantly moving. Using technical equipment, we collected two types of data to record the plants' responses (movement, electrical discharge):

Firstly, we collected video data by pointing cameras at a plant in a vegetable patch (Illustration 1) and simultaneously at the eurythmy dancer (Illustration 2) during an eurythmy dance performance.

Secondly, we tracked electric discharge by recording the voltage difference (i.e., electric discharge) from the leaves of the respective plants (Illustration 3) to the roots with a Plant SpikerBox, a voltage meter (Marzullo 2012). This allows us to measure the electrical potential difference between plant and soil.

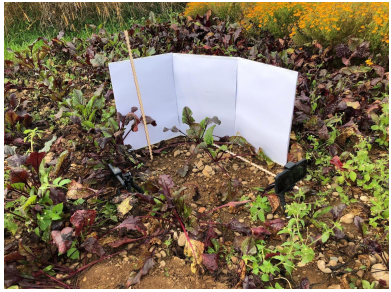 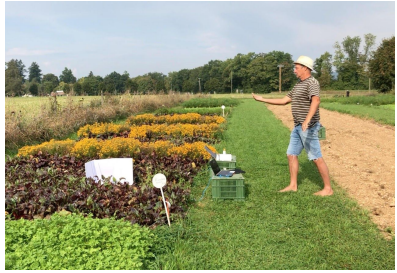 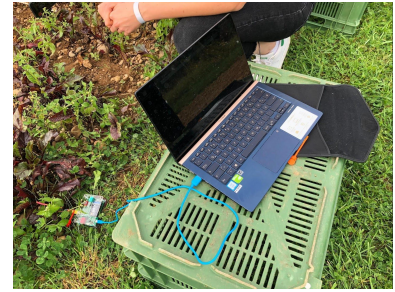

1: Recording of plant reactions  2: Eurythmy dancer in motion  3 Electric potential measurements

Next, we processed the data from our tracking devices. For the camera recordings of the plants, we firstly identified points of interest by applying the Shi-Tomasi Corner Detection method (Shi/Tomasi 1994). Using the Lucas-Kanade optical flow algorithm (Lucas/Kanade 1981), these points were tracked throughout the series of video frames to create a data series of vertical movement (Howse/Minichino 2020). For the camera recordings of the dancer, we used the same methods as for the plant recording. We tracked the dancer's hand movement, the strongest moving body part when conducting a eurythmy dance. In order to deal with the high number of outliers and noise caused by background and foreground disturbances as well as by the disappearance of the hands in front of the body, we did not produce continuous movement data, but distinguished between active dance and rest position. Therefore, we measured movement in a binary way (0: movement is not present, 1: movement is present) within defined bounding boxes next to the dancer's hands.

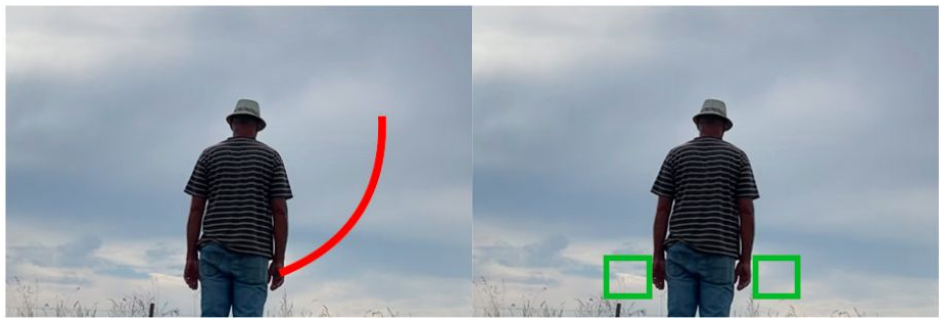

Display of motion recording with bounding boxes

For the electrical discharge data recorded by the SpikerBoxes, we extracted 20 Mel-Frequency Cepstral Coefficients (MFCCs) (Sigurðsson 2006) that split the complex electrical signal into a spectrum of different frequencies (Tahon/Devillers 2016). This is to filter out the majority of noise in the data via frequency decomposition (Sriram et al. 2005, Stanković et al. 2018). Next, the amplitudes (y-values) of the oscillations were correlated for different rolling windows for the two datasets extracted from the SpikerBox measurements. By further decomposing the amplitudes into individual frequencies, a considerable part of the data noise (caused e.g. by the laptops needed for measuring, field workers on adjacent fields, etc.) could be filtered out. The window length analysis revealed an optimal length of 21 seconds, resulting in the best compromise between sample size and correlation values (Ding et al. 2000, Yu 2015).

In our test series, we distinguished between plants regularly exposed to eurythmy in the past months and plants that had never been exposed to eurythmy before. We controlled the distance of the dancer to the plant by setting a defined point for conducting the dance. We kept the distance between the dancer and the measured plants consistently at 1m. Then, dances were conducted on average for 1.5 minutes per plant bed. The table below summarizes our experiments. In the next section, we present our data analysis and the corresponding results.

| Experiment | 1 | 2 | 3 | 4 | 5 | 6 | 7 | 8 | 9 | 10 |
|---|---|---|---|---|---|---|---|---|---|---|
| Plant | beetroot | beetroot | tomatoes | beetroot * | lettuce (first) | beetroot (first) | tomatoes (first) | beetroot | beetroot | tomatoes |
| Dance | Planets EU | Upper Sun | Mercury | Mercury | control, CCUL | control, CCUL | control, CCUL | Planets EU | Mercury | Mercury |

Table of experiments performed

\* *Note in table*: For subsequent analysis we discarded some measurement series due to measurement errors, equipment malfunction, and external influences.

## 4. Results

One explanation for why eurythmy has an effect on plants comes from anthroposophical perceptions of forces as energies that are a part of all biological movement-processes (Steiner 1924, Wachsmuth 1932). Our results seek to measure these perceptions of forces in three individual analyses.

### Analysis 1: Comparison of SpikerBox 1 und SpikerBox 2 Data Series

If the plant movement and discharge are really correlated to human movement, the two SpikerBoxes need to record similar signals. For each experiment, we therefore correlated the MFCCs of the two SpikerBoxes voltage time series. Comparing the MFCCs of the two SpikerBox voltage time series for beetroot, we found that specific MFCCs were indeed similar. The average correlation was formed over selected lower order MFCCs, as for all experiments in this paper. While higher order MFCCs carry specific details about the composition of the signal, lower order MFCCs capture the signal's general spectral shape (Peng et al. 2019). Hereafter, and in the further analyses, the given n represents the number of different plants measured.

| Beetroot Experiment# | Correlation |
|---|---|
| 01 | 0.588** |
| 02 | 0.464 |
| 06 | 0.467 |
| 08 | 0.389 |
| 09 | 0.639 |
| *Average* | **0.509** |

Results Analysis 1, Beetroot (n=12)

High correlations over almost the entire range of extracted MFCCs were found for the lettuce, which was danced at for the first time.

| Lettuce Experiment# | Correlation |
|---|---|
| 05 | 0.703** |

Results Analysis 1, Lettuce (n=2)

The lowest correlations across all experiments were found for tomatoes. It was found that the electrical discharge and the dancer's movement correlated for only higher frequencies of the direct cosine-transformed electrical signal.

| Tomatoes Experiment# | Correlation |
|---|---|
| 03 | 0.247 |
| 07 | 0.411 |
| 10 | 0.431 |
| *Average* | **0.363** |

Results Analysis 1, Tomatoes (n=6)

This analysis was conducted in order to investigate if changes in the environment have a measurable influence on the plants.

## Analysis 2: Comparison of electric discharge with movements of the dancer

We correlated the movements of the dancer with the electrical discharge of the plants. This analysis was conducted in order to investigate whether the eurythmy dancer himself has a direct effect on the plants. Measured effects would be in line with the etheric forces introduced by Rudolf Steiner (1924) and Keats (1999). The dancer's movement was coded with binary encoding for resting and active hands. The following table shows the average correlation coefficient over the lower order MFCCs for selected experiments.

|  | Left Hand | Right Hand |
|---|---|---|
| Experiment | ∅ Corr Coefficient | ∅ Corr Coefficient |
| 1 | 0.6034*** | 0.3950*** |
| 2 | -0.1294 | -0.1294 |
| 4 | 0.2997** | -0.4673*** |
| 6 | 0.2392*** | 0.2392*** |
| 8 | -0.5496*** | -0.3838*** |
| 9 | -0.2163 | -0.2846 |
| *Average* | **0.0412** | **-0.1051** |

Results Analysis 2 (n=12)

## Analysis 3: Comparison of electrical discharge of vegetables of the control group with electrical discharge of vegetables of the experimental group

Thirdly, we investigated if regularly danced plants react differently to eurythmy than plants that are exposed to dancing for the first time. We compared the electrical discharge for regularly danced and first time danced plants.

| Control | Correlation |
|---|---|
| 05 - Lettuce | 0.380 |
| 06 - Beetroot | 0.381* |
| 07- Tomatoes | 0.652** |
| *Average* | **0.471** |

Results Analysis 3, Control Group (n=6)

MFCCs that are considering the overall spectral shape are more meaningful for regularly danced plants. Furthermore, the dancing exerts little effect on the measurements of the two voltage meters.

| Experimental | Correlation |
|---|---|
| 01 - Beetroot | 0.558* |
| 02 - Beetroot | 0.398 |
| 03- Tomatoes | 0.196 |
| 08 - Beetroot | 0.172 |
| 09 - Beetroot | 0.548 |
| 10 - Tomatoes | 0.353 |
| Average | **0.371** |

Results Analysis 3, Experimental Group (n=12)

The last analysis reiterates that danced vegetables behave differently than those that were never danced before. Correlations in the control group (0.47) are higher than for long-term danced plants that were exposed to eurythmy (0.37).

## 5. Discussion

By conducting three analyses, we addressed our initial research question: *How do plants respond when being exposed to eurythmy dancing in terms of electrical discharge and movement?*

From analysis one, we conclude that immediate environmental changes (dancing or otherwise) have a measurable effect on plants in a biodynamic environment. As the two SpikerBoxes attached at the same time to two different plants in a plant patch show correlation in terms of electrical discharge, an anthroposophic researcher may say that *different plants experience the same etheric energies during a eurythmy dance performance*.

In analysis two, we found weak indications that plants may be influenced by eurythmy dances since there is a correlation between the plants and the dancer. In anthroposophic words, *it seems that the plant might feel the etheric energies of the dancer.*

In analysis three, we observe that plants react less to eurythmy dancing when being frequently exposed to it because the experimental group shows lower correlations than the control group.

Through an anthroposophic lens, *we speculate that plants become less sensitive to eurythmy when they are exposed to dances frequently while those that are exposed to eurythmy for the first time respond the strongest.*

The possibility that etheric energies are a link between eurythmy dancing and plant response (movement, electrical discharge) (Steiner 1924, Keats 1999) is supported by our experiments. While the aim of most anthroposophic experiments in the field (e.g., Retallack 1973, Ekici et al. 2007, Singh 1962) is to show what specific effects human stimulations have on plant movement, our motive was not simply to show the effect, but rather to measure how eurythmy influences plants in two dimensions (movement, electrical discharge). The results of our analysis furthermore indicate that the measurement method is effective.

# 6. Limitations

Our research has different limitations. These are, among others, external influences caused by vibrations, weather patterns as well as a potential influence by researchers: Firstly, changes in the electrical potential could have been caused by the small, slow, and non-measurable vibrations of the dancer transmitted through the earth. By keeping a distance of one meter we tried to minimize the possible influence of such vibrations. Weather conditions such as strong winds and sun could also influence the measurements. To control for this, a day with low wind speed (<5km/h) was chosen for the experiments. Additionally, a windbreak was placed around the plants. Furthermore, experiments were conducted several times that day to take into account deviations due to the position of the sun.

Other external movements could also have had an influence on the plants. To avoid this, no measurements were taken when possible distractions, for instance passing tractors, were present. Moreover, the researchers kept a distance of at least three meters during data collection and allowed the plants to rest for at least 30 minutes after dancing.

Finally, our sample size is small, many more measurements need to be made to collect larger samples allowing us to obtain more significant results.

# 7. Conclusion

The results described in this paper indicate possible effects of eurythmy dancing on the response measured through leaf movement, and electrical discharge of plants. The results of this exploratory research are only indicative and directional as the number of experiment samples is small. What Steiner (1924) describes as etheric life energies might contribute to an explanation for eurythmy's possible influence on plants. Our results need to be corroborated by further experiments on plants at biodynamic research facilities including additional environmental factors such as microclimate (e.g., humidity, particular matter).